\documentclass[a4paper]{article}
\usepackage{wstyle,amsmath,epsfig}

\title{When correlations matter - response of dynamical networks to small perturbations}
\name{Thimo Rohlf$^{1,2}$, Natali Gulbahce$^{3,4}$ and Christof Teuscher$^5$}
\address{$ ^1$Santa Fe Institute, 1399 Hyde Park Road, Santa Fe, NM 87501, USA \\
$ ^2$Max-Planck-Institute for Mathematics in the Sciences, 
Inselstrasse 22, D-04103 Leipzig, Germany\\ 
$ ^3$ Center for Complex Network Research, Northeastern University, Boston,
MA 02115, USA\\
$^4$ Center for Cancer Systems Biology, Dana Farber Cancer Institute,
Boston, MA, 02215, USA\\
$ ^5$ Los Alamos National Laboratory, 
CCS-3, MS B256, Los Alamos, NM 87545, USA\\
rohlf@santafe.edu }
\begin{document}

\maketitle
\begin{abstract}
We systematically study and compare damage spreading
for random Boolean and threshold networks under small external
perturbations (damage), 
a problem which is relevant to many biological networks.
We identify a new characteristic
connectivity $K_s$, at which the average number of damaged
nodes after a large number of dynamical updates is independent
of the total number of nodes $N$. We estimate the critical
connectivity for finite $N$ and show that
it systematically deviates from the annealed approximation. 
Extending the approach followed in a previous study \cite{Rohlf2007a},
we present new results indicating that internal
dynamical correlations tend to increase not only the probability
for small, but also for very large damage events,
leading to a broad, fat-tailed distribution of damage sizes.
These findings indicate that the descriptive and predictive
value of averaged order parameters for finite size networks
- even for biologically highly relevant sizes up to several thousand nodes - is limited.
\end{abstract}
\section{INTRODUCTION}
\label{sec:intro}
Random Boolean networks (RBN) were originally introduced as simplified
models of gene regulation \cite{Kauffman1969a,Kauffman1993}. In the 
limit of large system sizes, they exhibit a dynamical
order-disorder transition at a critical wiring density $K_c$
\cite{Derrida1986a}; similar observations were made for sparsely
connected random threshold (neural) networks (RTN)
\cite{Kuerten1988b,Rohlf2002}.  For a finite system size
$N$, the dynamics of both systems converge to periodic attractors
after a finite number of updates. At $K_c$, the phase space structure in
terms of attractor periods \cite{Albert2000}, the number of
different attractors \cite{Samuelsson2003} and the distribution of
basins of attraction \cite{Bastolla1998} is complex.
Furthermore, critical networks exhibit many
properties reminiscent of biological networks,
leading to the idea $K_c$ might be an "attractor of
evolution" \cite{Kauffman1993}.

To ensure proper function, regulatory networks in living cells
have to be robust (insensitive) against external perturbations.
In terms of RBN/RTN dynamics, perturbations can disrupt the generic
dynamical state (fixed point or periodic attractor) of the network,
and hence are referred to as "damage"; this type of study
has been applied, for example, to the perturbation of gene expression
patterns in a cell due to mutations \cite{Ramoe2006}.

Mean-field techniques as, for example, the {\em annealed approximation} (AA)
introduced by
Derrida and Pomeau \cite{Derrida1986a}, allow for an analytical treatment of
damage spreading and exact determination of the critical connectivity $K_c$
under various constraints \cite{Sole1995}.  It has been shown
that local rewiring rules coupled to mean-field-like order parameters of the
dynamics can drive both RBN and RTN to self-organized criticality
\cite{Bornholdt2000a,Liu2006}.

Studies of RBN/RTN dynamics based on the AA usually implicitly assume
that, at least for large $N$, {\em principal} properties
of damage spreading should not depend on the initial perturbation size.
For example, the determination of $K_c$ using a one-bit initial
perturbation (sparse percolation limit), or an initial perturbation size increasing with $N$
should yield the same value for large $N$, since it is assumed that
correlations can be neglected in this limit by averaging
over a large number of different random network realizations.
In this paper, we extend results of a previous study \cite{Rohlf2007a}
and present the following findings that are, at least in part, in clear contradiction
to these assumptions:
\begin{itemize}
\item In section 3.1, we identify a new characteristic point $K_s < K_c$, 
where the expectation value of the number of damaged nodes after large number
of dynamical updates is independent of $N$.
\item By the definition of marginal damage spreading, we
estimate the critical connectivity $K_c(N)$ for
finite $N$, and present
evidence that, even in the large $N$ limit, for small
initial perturbations $K_c$
systematically deviates from the predictions of the AA (section 3.2).
\item In section 3.3, we present new results proving that, slightly below $K_c$, starting from
random initial conditions, the AA holds only for small times $t$,
indicating that after passing transient dynamics
inherent correlations considerably affect damage propagation.
\item Last, we show that vanishing, as well as large damage events
are overrepresented in damage size statistics, leading to highly skewed
distributions, which are poorly characterized by averages (section 3.4).

\end{itemize}

\section{DYNAMICS}
\label{sec:dynamics}

\subsection{Random Boolean Networks}
\label{ssec:rbn}
A RBN is a discrete dynamical system composed of $N$ automata. 
Each automaton is a Boolean variable with two possible states: 
$\{0,1\}$, and the dynamics is such that
\begin{equation}
{\bf F}:\{0,1\}^N\mapsto \{0,1\}^N, 
\label{globalmap}
\end{equation} 
where ${\bf F}=(f_1,...,f_i,...,f_N)$, 
and each $f_i$ is represented by a look-up table of $K_i$ inputs
randomly chosen from the set of $N$ automata. Initially, $K_i$ 
neighbors and
a look-table are assigned to each automaton at random.

An automaton state $ \sigma_i^t \in \{0,1\}$ is updated using
its corresponding Boolean function:
\begin{equation}
\sigma_i^{t+1} = f_i(\sigma_{i_1}^t,\sigma_{i_2}^t, ... ,\sigma_{i_{K_i}}^t).
\label{update}
\end{equation}
We randomly initialize the states of the automata (initial 
condition of the RBN). The automata are updated synchronously using their
corresponding Boolean functions.

\subsection{Random Threshold Networks}
\label{ssec:titlestyle}
An RTN consists 
of $N$ randomly interconnected binary sites (spins) with states $\sigma_i=\pm1$.
For each site $i$, its state at time $t+1$ is a function of the inputs it receives 
from other spins at time $t$:
\begin{eqnarray} 
\sigma_i(t+1) = \mbox{sgn}\left( \sum_{j=1}^N c_{ij}\sigma_j(t) + h.    \right) 
\end{eqnarray}  
The $N$ network sites are updated synchronously. In the following
discussion the threshold parameter $h$ is set to zero. The interaction weights
$c_{ij}$ take discrete values $c_{ij} = +1$ or $-1$ with equal
probability. If $i$ does not receive signals from $j$, one has $c_{ij} = 0$.\\

\begin{figure}[t]
\centering
\includegraphics[width=8cm, height=3.5cm]{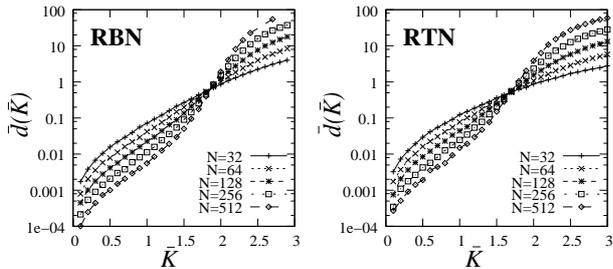}
\caption{\small Average Hamming distance (damage) $\bar{d}$ after 200 system
updates, averaged over 10000 randomly generated networks for each value of
$\bar{K}$, with 100 different random initial conditions and one-bit perturbed
neighbor configurations for each network. For both RBN and RTN, all curves for
different $N$ approximately intersect in a characteristic point $K_s$.}
\label{fig1}
\end{figure}

\begin{figure}[t]
\centering
\includegraphics[width=8cm, height=3.5cm]{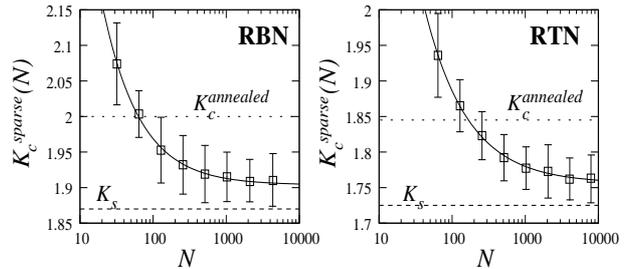}
\caption{\small The critical connectivity $K_c^{sparse}(N)$ in the SP limit as
a function of $N$.
Curves are
power-law fits according to Eq. (\ref{kcscale_eqn}), straight dashed lines mark $K_c^{annealed}$ and $K_s$ for
RBN and RTN, respectively.}
\label{fig2}
\end{figure}

\section{Results}

\subsection{Scaling}
We first study the expectation value $\bar{d}$ of damage,
quantified by the Hamming distance of two different system configurations,
after a large number $T$ of system updates.  
Fig. \ref{fig1} shows $\bar{d}$ as a function of the average connectivity
$\bar{K}$ for different network sizes $N$ by using a random ensemble
for statistics.  For both RBN and RTN, the observed functional behavior strongly
suggests that the curves approximately intersect at a common point $(K_s, d_s)$,
where the observed Hamming distance for large $t$ is independent of the system
size $N$. 

We verified this finding quantitatively by using finite-size-scaling methods \cite{Rohlf2007a}.
In particular, one can show that $\bar{d}$ as a function of $N$ and $\bar{K}$
obeys the following scaling ansatz:
\begin{equation}
\bar{d}(\bar{K}, N) = a(\bar{K})\cdot N^{\gamma(\bar{K})} + d_0(\bar{K}),\,-1 \le \gamma \le 1. \label{fs_eqn}
\end{equation}
It is straight-forward to show that $\gamma \to -1$ for small $\bar{K} \to 0$, and that $\gamma \to 1$
for densely connected networks above the percolation transition ($\bar{K} > K_c$). Evidently, this
implies that at some characteristic connectivity $K_s$, there has to be a transition from negative to
positive $\gamma$ values, with $\gamma(K_s) \approx 0$. It is a very interesting question
whether $K_s$ coincides with $K_c$, or if it is different from $K_c$ for large $N$.
For a precise numerical determination of $K_s$, one can make use of the fact that $\bar{d}$
exhibits an exponential dependence near $K_c$:
\begin{equation}
\bar{d}(\bar{K}, N) \approx c_1(N)\,\exp{[c_2(N)\,N^{\alpha}\,\bar{K}]} \label{exp_eqn}
\end{equation}
with $\alpha \approx 0.42$. High-accuracy fits of this dependence (with $c_1$  and $c_2$
as adjustable parameters) in the interval $1.6 \le \bar{K} \le 2.1$ yield 
\begin{equation}
(K_s^{RBN}, d_s^{RBN}) = (1.875 \pm 0.05, 0.62 \pm 0.05)
\end{equation}
for RBN and, correspondingly,
\begin{equation}
(K_s^{RTN}, d_s^{RTN}) = (1.729 \pm 0.045, 0.51 \pm 0.04)
\end{equation}
for RTN. We verified these findings up to $N=16384$, waiting $T=5000$ updates for the dynamics
to relax; for even larger $N$,
simulations become intractable due to exponentially increasing relaxation times. Evidently,
we tend to miss large damage events since they need the most time to develop. 
Facing this unavoidable {\em biased undersampling} of large avalanches, one can argue
that the {\em true} values of $K_s$ are probably even lower than our measured values.
From this evidence, and also from more refined scaling arguments \cite{Rohlf2007a}, we conclude
that $K_s$ is distinct from $K_c$ in the limit of large $N$.

\subsection{Deviations of $K_c$ from the annealed approximation}
Interestingly, $K_s$ is close to, but distinct from the critical
connectivities $K_c^{RBN} =2$ and $K_c^{RTN} =1.845$, as predicted by
the AA. Since in this study we consider the limit of
very weak initial perturbations which is usually not covered
in theoretical studies of RBN/RTN dynamics, we 
now have to consider the possibility that $K_c$ itself may deviate
from the prediction of the AA. An
intuitive definition of criticality for finite $N$ can be formulated
in terms of {\em marginal damage spreading}. If at time $t$ one bit is
flipped, one requires at time $t+1$ \cite{Sole1995,Rohlf2002}
\begin{equation}
\bar{d}(t+1) = \langle p_s\rangle(K_c) K_c = 1, 
\end{equation}
where $\langle p_s\rangle(\bar{K})$ is the average damage propagation
probability.
Fig. 2 shows $K_c^{sparse}(N)$, using the values $c_1(N)$ and $ c_2(N)$
obtained from numerical fits of Eq. (\ref{exp_eqn}) for both RBN and RTN. We find
that both systems, in a very good approximation, obey the scaling relationship
\begin{equation}
K_c^{sparse}(N) \approx b\cdot N^{-\delta} + K_c^{\infty}\label{kcscale_eqn}
\end{equation}
with $b = 3.27 \pm 0.79$, $\delta = 0.85 \pm 0.07$ and $K_c^{\infty} = 1.9082 \pm 0.008$ for RBN and $b = 3.853 \pm 0.76$,
$\delta = 0.736 \pm 0.05$ and $K_c^{\infty} = 1.7595 \pm 0.008$ for RTN.
Hence, in the limit $N \to\infty$, we can extrapolate
\begin{equation} K_c^{\infty, RBN} =
1.9082 \pm 0.008
\end{equation}
for RBN, and for RTN
 \begin{equation}
 K_c^{\infty, RTN} = 1.7595 \pm 0.008.
\end{equation}
Thus, for both RBN and RTN in the sparse percolation limit, we make the surprising observation that $K_c^{sparse}$ systematically 
deviates from $K_c^{annealed}$. While we find  $K_c^{sparse}(N) > K_c^{annealed}$ 
for small $N < 128$, for larger $N$ we observe a monotonic decay that approaches an 
asymptotic value considerably below $K_c^{annealed}$, suggesting that the observed
deviations from the AA also hold in the large $N$ limit. 
In the following two subsections, we will extend this analysis and discuss possible causes for these
deviations.

\begin{figure}[t]
\centering
\includegraphics[width=8cm, height=6.0cm]{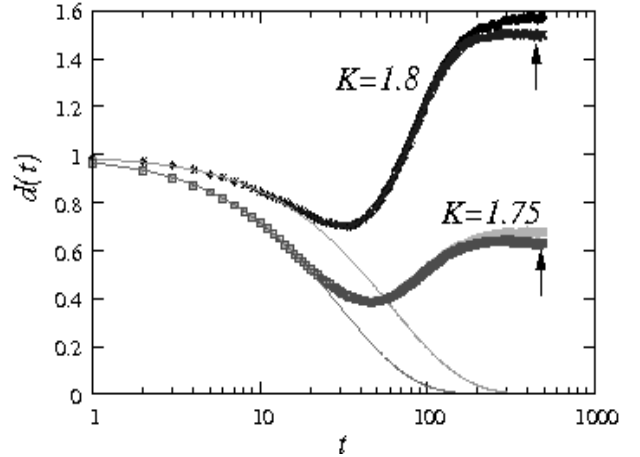}
\caption{\small Time-dependence of (average) damage propagation in RTN of size $N=4096$ just below $K_c$; damage $\bar{d}$ at time $t$ was
averaged over $10^5$ network realizations and $100$ different initial conditions 
(and the corresponding
neighbor states with one bit perturbed at random) at $t=0$
for each data point . Lined curves are the corresponding solutions of the
AA (Eq. (\ref{AA_time_eq})). For $t \ge 20$, pronounced deviations of simulation results from the AA are found,
in particular for $\bar{K} = 1.8$. Arrows indicate results with "corrected" statistics, i.e. without "pseudo-damage"
due to attractor phase lags.
}
\label{fig3}
\end{figure}

\subsection{Time dependence of $\bar{d}$}
Since we found systematic deviations from the AA for large $t$, it is interesting to ask
whether the AA still holds for small $t$, starting from random initial states. In particular,
one can derive the following recursive map for damage propagation at $t>0$ \cite{Rohlf2002}:
\begin{equation}
\bar{d}(t) = N\cdot \langle p_s\rangle(\bar{K})\cdot\left(1-e^{-\bar{K}\cdot \bar{d}(t-1)/N}\right), \label{AA_time_eq}
\end{equation}
where $\langle p_s\rangle(\bar{K})$ is the average probability that a link
propagates damage.
Let us now test this relationship in the interesting range $K_s \le \bar{K} \le K_c^{annealed}$
for ensembles of randomly generated networks (RTN with Poissonian degree-distribution),
with one-bit perturbations of randomly chosen initial conditions.
Figure 3 shows that, for small $t$, the dependence for $\bar{d}(t)$ found in numerical simulations obeys
this prediction very well. However, after an 
initial decrease of $\bar{d}(t)$, an {\em increase} above the initial damage size (i.e. supercritical
behavior) is found, in clear contradiction to the AA. This indicates that, after the system has passed
transient dynamics, inherent dynamical correlations considerably modify damage propagation (fractal structure
of attraction basins \cite{Bastolla1998}). One can also show that "pseudo-damage" events , i.e. cases where networks run on the same attractor,
but with a phase lag captured in a non-zero Hamming distance, do {\em not} substantially contribute to this effect
(arrows in Fig. 3). This proves that our results are very robust against changes in the way statistics is taken.

\begin{figure}[t]
\centering
\includegraphics[width=8cm, height=5.0cm]{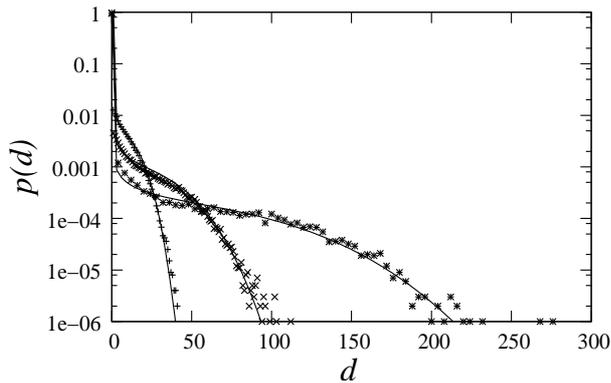}
\caption{\small Statistical distribution $p(d)$ of damage sizes for three different system sizes: $N=64$ (+), $N=256$ (x) and
$N=1024$ (*). Lined curves are solutions of Eq. (\ref{damdist_theo_eq}).}
\label{fig4}
\end{figure}

\subsection{Distribution of damage sizes}
Let us now go beyond averaged (mean-field) quantities and investigate detailed statistics of damage sizes.
For this purpose, for different $\bar{K}$ and $N$ ensembles of $Z_e$ random network realizations were created; for each network
realization, $Z_i$ random initial conditions $\vec{\sigma}$ (plus a neighbor state with one bit perturbed at random)
were tested, and statistics of damage sizes was taken after 1000 dynamical updates. Notice that we do {\em not} average damage
sizes for a given network realization, since this would again represent a kind of mean-field approximation.
Figure 4 shows that the resulting statistical distributions near $K_c$ are highly skewed, with more than $90\%$ events of vanishing damage size,
and a flat tail of large damage events which becomes more and more pronounced for increasing $N$. Similar problems have been
studied by Samuelsson and Socolar \cite{Samuelsson2006} for the number of {\em undamaged} nodes $u$ in the limit of {\em exhaustive} percolation. From
symmetry considerations, it follows that the probability distribution $p(d)$ of the number $d$ of {\em damaged} nodes in the limit of sparse percolation obeys a similar dependence as $u$ in the case of exhaustive percolation,
and hence
\begin{equation}
p(d) \approx a(N)\cdot \frac{\exp{[-\frac{1}{2}(d\cdot N^{-2/3})^3]}}{\sqrt{d\cdot N^{-2/3}}} \label{damdist_theo_eq},
\end{equation}
where $a(N)$ is a free parameter. One finds that the results of numerical simulations agree very well with this estimate
even for considerably small $N$ (Fig. 4). 
From the shape of these distributions, one recognizes that vanishing, as well as large damage events are much more probable
than expected from mean-field considerations. In part, this explains the deviations from the annealed approximation 
found for $\bar{d}$ near criticality (Fig. 3), and it also questions in how far averaged quantities deliver an informative description
of RBN/RTN dynamics for finite size $N$.

\section{DISCUSSION}
We showed that, for very weak (one-bit) perturbations of the initial states of RBN and RTN dynamics, the resulting damage
at later times exhibits a non-trivial scaling with network size $N$, and, near the critical order-disorder transition -
the so-called the 'edge of chaos' - considerable deviations from the annealed approximation. These deviations have escaped earlier studies,
since usually the {\em rescaled} damage $\bar{d}/N$ (or the overlap $1-\bar{d}/N$, respectively) was studied, and the thermodynamic
limit of large $N$  was considered. Our study indicates that there is a strong need for more refined studies of damage
propagation in RBN/RTN, that explicitly take into account dynamical correlations and the fractal structure of attraction basins \cite{Bastolla1998}.
One may expect that the situation is even more complex for networks with more realistic topologies. Even for simple random graphs,
as applied in this study, damage size distributions are highly skewed, questioning the descriptive and predictive value
of simple, averaged order parameters for this class of complex systems.


\bibliographystyle{plain}
\bibliography{refs}

\end{document}